\documentclass{optica-article}

\journal{opticajournal} 

\articletype{Research Article}

\usepackage{lineno}
\usepackage{amsmath}
\usepackage{esvect}
\linenumbers 

\begin{document}
\nolinenumbers

\title{Single shot multi-time frame imaging of plasmas with a frequency tagged GHz pulse train}

\author{Justin Twardowski,\authormark{1,*} Conrad Kuz,\authormark{2} Mohamed Yaseen Noor,\authormark{1} Andy Lee,\authormark{2} Jason Leicht,\authormark{3} Mikhail Slipchenko,\authormark{3} and Enam Chowdhury\authormark{1,4,2}}

\address{\authormark{1} Department of Material Science and Engineering, The Ohio State University, Columbus, OH 43210, USA\\
\authormark{2} Department of Physics, The Ohio State University, Columbus, OH 43210, USA\\ 
\authormark{3} Spectral Energies LLC, Beavercreek, OH 45430, USA\\
\authormark{4} Department of Electrical and Computer Engineering, The Ohio State University, Columbus, OH 43210, USA\\}
\email{\authormark{*}twardowski.6@osu.edu} 

\begin{abstract*} 
Studying plasma dynamics is crucial for understanding processes like inertial confinement fusion, material damage, and shockwave formation from intense laser or current interactions. While pump-probe methods are standard for capturing these dynamics, single-shot experiments using high-power, low-repetition-rate systems with custom targets are challenging. We present a novel imaging technique using a synchronized GHz-rate spectrally tagged probe laser to capture multiple time-resolved snapshots from a single pump event. Diffraction optics spatially separate the pulses, yielding nanosecond-resolved images. Experiments with tantalum and borosilicate glass show shockwave velocities consistent with literature and Sedov scaling, validating the method’s accuracy and utility in plasma diagnostics.
\end{abstract*}

\section{Introduction}
The pump-probe method, where a probe source pulse synchronized with the pump pulse illuminates the interaction region driven by the pump pulse at specific delays, is extremely well-suited to capture various dynamical processes. In a small system with a reproducible interaction region, this is relatively straight-forward to implement, as one can change the relative time delay between the pump and the probe and repeat the experiments to capture the time-evolution of the interaction. Pump-probe experiments have been used widely, from observing capsule implosion in the inertical confinement fusion process \cite{Rygg2014} to various laser matter interactions, like the changing electron density in molecules \cite{stock1992detection}, excited carrier recombination times \cite{terada2010real}, melting dynamics in solids \cite{reitze1989femtosecond}, ablation processes in single and multilayer dielectrics \cite{talisa2018few,talisa2020comparison}, and shock wave propagation \cite{hattori2020investigation}. In particular, studying shock waves or propagating disturbances that travel faster than the speed of sound in a material—is critical for understanding both material processing dynamics and intrinsic material properties. In the ablation of silica glass, three stress waves are visible, the first being attributed to thermal expansion of the target and the latter two being due to the pressure of the laser \cite{hu2010generation}; while, in high repetition rate laser processing, repeated pressure waves lead to tensile stress and eventually cracks around the processed area \cite{ito2019dynamics}. In addition, the measurement of photo-acoustic transduction in metals is carried out by by generating shockwaves \cite{kaleris2023efficient}. Observing dynamics with temporal resolution is often fundamental in understanding these physical and electronic processes.

High energy high intensity lasers, on the order of 10\textsuperscript{15} - 10\textsuperscript{18}~W/cm\textsuperscript{2} and beyond can access high energy density conditions of matter \cite{Park2021}, inertial confinement fusion \cite{Hurricane2014} to reaching the relativistic regime, wherein electrons on the surface of a target are accelerated close to the speed of light \cite{salamin2006relativistic, palaniyappan2012dynamics}. A process that can occur in relativistic laser plasma interaction (RLPI) is relativistic transparency, where more laser energy can be coupled into the target \cite{palaniyappan2012dynamics}, which could be turned into more high energy X-rays for radiography \cite{sefkow2011efficiency,courtois2013characterisation}. High Z targets, like tantalum, are used to generate Bremsstrahlung radiation, and coating in a lower Z material may increase the laser-target coupling \cite{strehlow2024mev}. A coating can improve laser-target coupling depending on the laser contrast and duration, and the coating density and thickness \cite{strehlow2024mev,yin2024advances}.

While innovating pump-probe techniques exist for high repetition rate RLPI experiments \cite{Feister2014, Patnaik:25}, many high intensity lasers with high energy/pulse often have low repetition rates, on the order of 1 Hz or less, which can slow data acquisition. In addition to low repetition rate, high-intensity laser-plasma interactions often include shot-to-shot instabilities because of nonuniform plasma density, changing refractive indices and the laser intensity \cite{umstadter2003relativistic}. Due to low repetition rate and plasma instability, it is of critical importance to be able to characterize high intensity laser-plasma interaction in a single shot, and significant efforts are invested in large scale facilities like at the National Ignition Facility at the Lawrence Livermore National Laboratory \cite{Khan2021, Hill2021} or at the Z-backlighter facility \cite{Rambo2019} at the Sandia National Laboratory to develop appropriate probe diagnostics. Additionally, specially manufactured structured targets can be expensive and time-consuming to produce. The ability to collect several probe images in a single pump shot, on an expensive target, would provide more information about these instabilities and decrease time required for data collection. A single shot imaging scheme \cite{Rambo2019} may also assist in elucidating other unstable plasma behavior, such as plasma generated by x pinch, z pinch, or dense plasma focus - which have uses in radiography, x-ray lithography, and neutron generation \cite{umstadter2003relativistic,pikuz2001spatial,krishnan2012dense}.

In this work, we image shockwaves generated by both femtosecond and nanosecond long pump pulses on bare tantalum and plastic-coated tantalum with a novel imaging and probe scheme. The shockwaves are probed using an independent probe laser system that generates a GHz-rate pulse train synchronized with the pump pulse, whose individual pulses are spectrally tagged and temporally separated by 1~ns from the previous and sub-sequent pulses. Each pulse is spectrally offset by $\sim$1~nm, enabling multi time-frame imaging of the LPI dynamics within a single pump-probe acquisition event. 

\section{Methods}
\begin{figure}[h]
\centering
\fbox{\includegraphics[height=.5\textwidth]{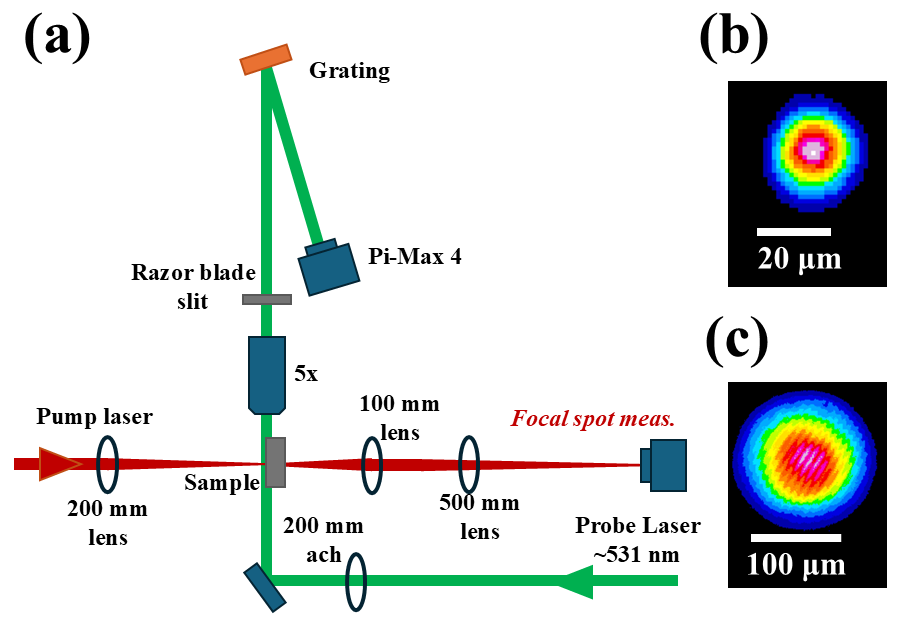}}
\caption{(a) The experimental setup for orthogonal imaging. With either laser, the pump is focused with a 200~mm lens. After the 5x objective, the beam goes out of the page from the grating. Focal spots from the (b) 780~nm, 40~fs pump and the (c) 527~nm, 20.4~ns pump.}
\label{fig:Setup}
\end{figure}

A 780 nm, 40 fs Ti:Sapphire home-built laser was normally incident on a $5\times 20\times1$~mm strip of bare tantalum or 12 um thick (AZ nLoF 2070) photoresist-coated Ta. Peak intensities ranged from 1.90 - 6.92 $\times$ 10\textsuperscript{15}~W/cm\textsuperscript{2} with a 1/e\textsuperscript{2} diameter of 30.5~$\mu$m. A Northrop Grumman (NG) Gigashot T Laser Model GST-012-QMF-0100 with 527~nm, 20~ns was also used to excite the same tantalum targets, with peak intensities from 1.40 - 2.33 $\times$ 10\textsuperscript{9}~W/cm\textsuperscript{2} with a diameter of 178.0~$\mu$m. A comprehensive diagram showing the experimental setup is detailed in Fig. \ref{fig:Setup}(a). Focal spots from both the femtosecond and the nanosecond pump are shown in Fig. \ref{fig:Setup}(b)-(c). A higher intensity was possible with the nanosecond laser; however, pump flash centered around 527~nm could not be entirely negated.

\begin{figure}[h]
\centering
\fbox{\includegraphics[height=.34\textwidth]{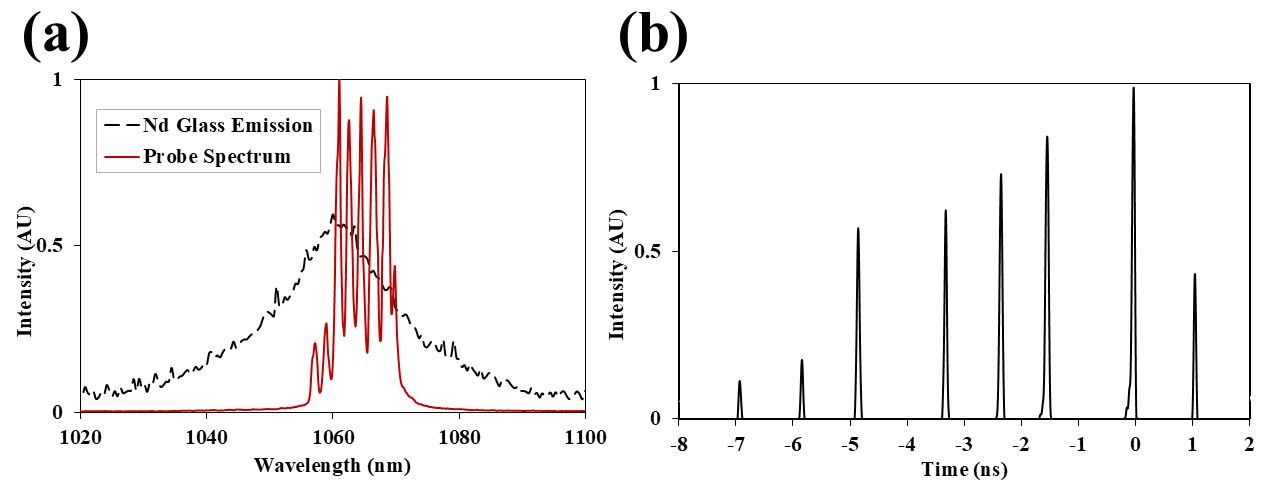}}
\caption{The probe’s spectral and temporal characterization. (a) The probe spectrum, with the freespace Nd:Glass gain bandwidth overlayed, illustrating the pump amplifier’s usable amplification window (and subsequent 532~nm spectrum).  (b) The temporal impulse response recorded on a Tektronix 23~GHz oscilloscope with high speed 30~GHz photodiode. Together, these measurements confirm that the probe train maintains precise spectral and temporal separation.}
\label{fig:Spectral}
\end{figure}

\begin{figure}[h]
\centering
\fbox{\includegraphics[height=.45\textwidth]{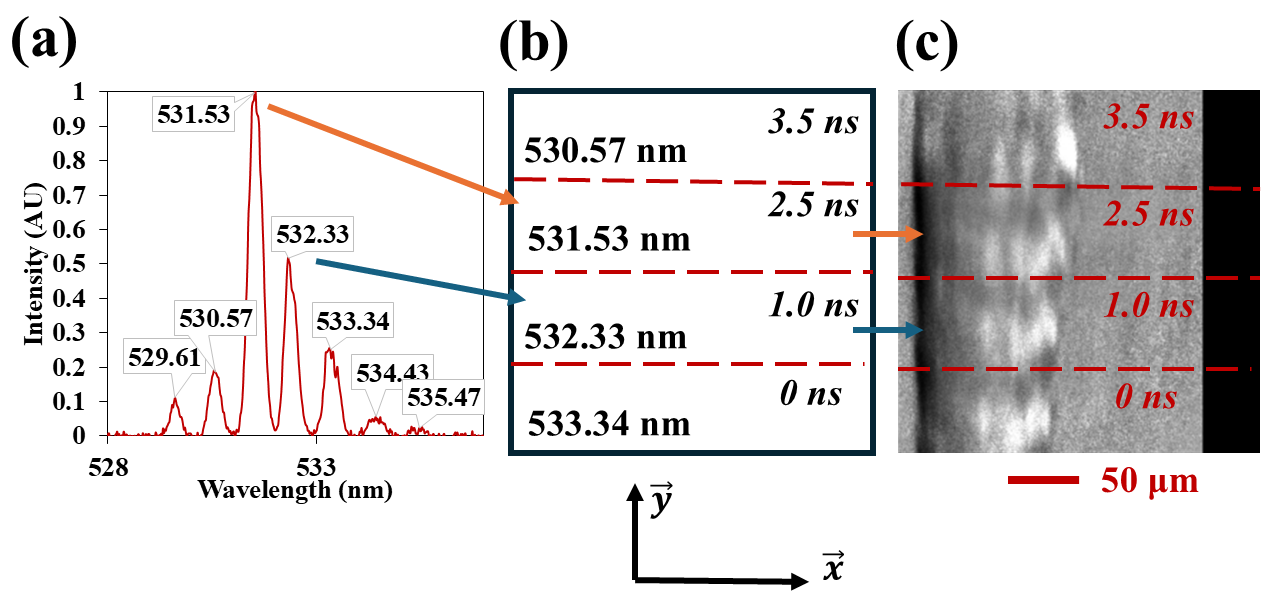}}
\caption{(a) Spectrum of the probe pulse. The overall bandwidth of the laser is split into seven pulses, and then separately filtered with an angled-tuned narrow bandpass filter. After these spectrally distinct pulses are individually delayed. (b) Overview of obtained probe image. Due to the diffraction grating, the image is stretched or distorted in $\protect\vv{y}$, but unaffected in $\protect\vv{x}$. (c) Actual background subtracted pump-probe image, showcasing the four channels. The laser is incident from -$\protect\vv{x}$, and the shockwaves propagates in $\protect\vv{x}$.}
\label{fig:Spectrum}
\end{figure}

The probe arm consists of a GHz-rate laser from Spectral Energies. A single 4~ps, 100~pJ, 1064~nm broadband pulse is first stretched to 1.1~ns in a fiber stretcher and amplified by an ytterbium-doped fiber amplifier. The stretched pulse is then routed through a fiber splitter, with each channel passing through a tunable 1~nm-bandwidth bandpass filter to yield $\sim$2~nm spectral spacing between subsequent pulses. Temporal separation of $\sim$1~ns is then introduced through fiber delay lines. The time and wavelength encoded pulse replicas are recombined, the fiber spectral and temporal profiles are shown in Fig. \ref{fig:Spectral} and the total energy in each pulse packet (all 8 pulses) is ~0.015 uJ.  These pulses are transferred to free space, where they undergo amplification in a neodymium-doped silicate glass (Q246) booster amplifier to a total energy of  0.157 mJ. Finally, the amplified pulses are frequency-doubled in a $\beta$-barium borate (BBO) crystal with each pulse packet having 70~nJ centered around 532~nm. The resulting pulse spectrum is presented in Fig. \ref{fig:Spectrum}(a).

The probe is oriented orthogonally to the sample, and imaged with a $5\times$ microscope objective. A grating with 1800~l/mm is used in the imaging line to separate the probes vertically, or out of the page. The camera (Pi-Max4 1024f) then images several distinct timestamps in the sample's development from a single pump shot. The captured image is subdivided into four images, as in Fig. \ref{fig:Spectrum}(b). While seven individual probes were available, only four were used as they had the greatest intensity. The image is slightly smeared in one direction but is unaffected in the other. The spatial resolution in the horizontal direction is 2.19~$\mu$m with a 253~$\mu$m field of view. A diffraction grating with a line spacing of 1800~l/mm was necessary to separate the channels; 0.13\textdegree~separation was experienced by the channels, with their peaks 1~nm apart. Before the grating a razor blade slit was implemented to cut the image into a horizontal band - this is necessary otherwise the full image would overlap with other channels.

Additionally, the Ti:Sapphire pump was used to excite a glass microscope coverslip whilst imaging in transmission. In this configuration, the pump was at 45\textdegree~angle of incidence. The probe was sent through the sample, and was also imaged in transmission. Peak intensity on the target ranged from 1.93 - 3.60 $\times$ 10\textsuperscript{15}~W/cm\textsuperscript{2}. In all the aforementioned trials, the timing was established using a photodiode (DET025AFC) that detected the pump flash in either the reflected or transmitted direction. The oscilloscope used to monitor timing was a InfiniiVision MSOX4154A 1.5~GHz. An electronic shutter, synchronized via an external timing box, controlled both pump and probe allowing straightforward adjustment of probe delay.

\section{Results}
\subsection{Bare Ta and plastic on Ta}
\begin{figure}[hbt!]
\centering
\fbox{\includegraphics[height=.35\textwidth]{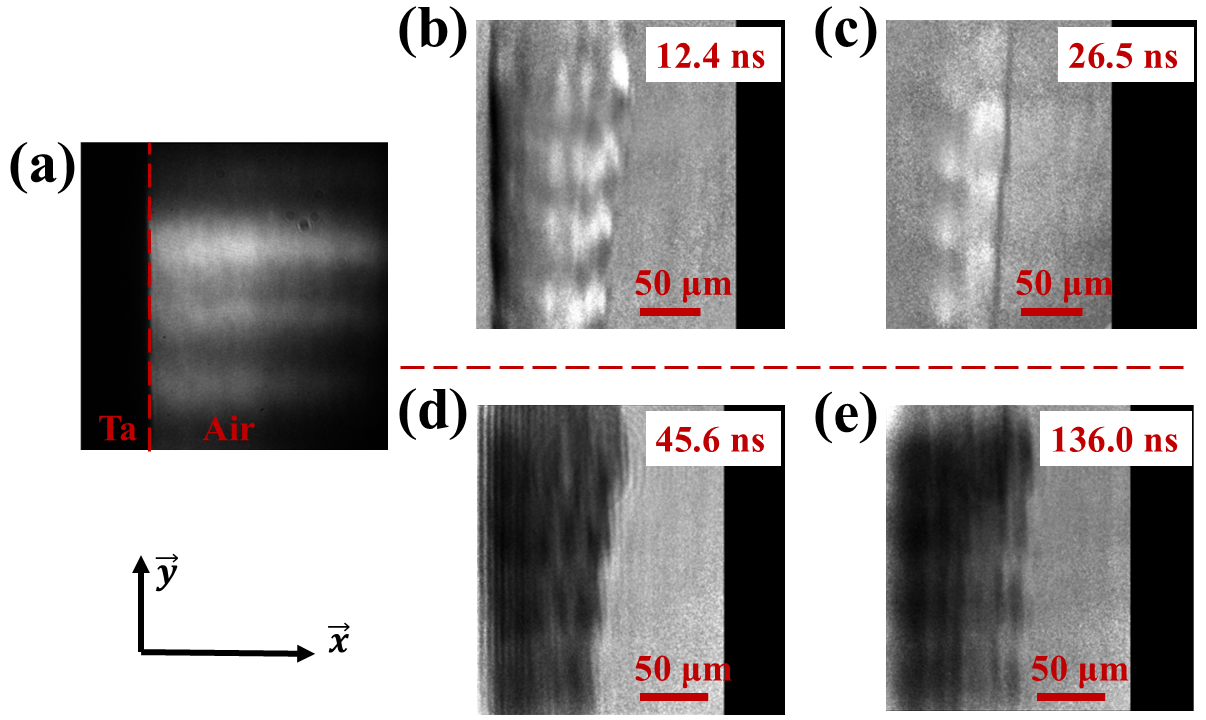}}
\caption{(a) Before image of the target, with the tantalum on the left and air on the right. The four channels are visible as horizontal bands. The during images are obtained through background subtraction, and a shift in -$\protect\vv{x}$ to have the shockwave begin on the left side of the image. The time the first channel of the probe arrived after the pump is detailed in the upper right-hand side of each during image. The 780~nm, 40~fs pump was used on (b) bare Ta  and (c) plastic-coated Ta. The 527~nm, 20.4~ns pump was incident on (d) bare Ta and (e) plastic-coated Ta .}
\label{fig:Actual}
\end{figure}

Representative images of shockwaves generated on bare Ta and plastic coated Ta are illustrated in Fig. \ref{fig:Actual}. The laser is incident from the right of the image, and the shockwaves begin on the left side of the image, and propagate in +$\protect\vv{x}$.

Shockwaves derived from the femtosecond laser originate, Fig. \ref{fig:Actual}(b)-(c) much earlier from the sample, and progress much more quickly across the field of view. The peak intensity used is 5.51 $\times$ 10\textsuperscript{15}~W/cm\textsuperscript{2} on bare Ta, Fig. \ref{fig:Actual}(a) and 6.53 $\times$ 10\textsuperscript{15}~W/cm\textsuperscript{2} on the photoresist-coated Ta, Fig. \ref{fig:Actual}. Between the two, as each channel of the probe is delayed by 1~ns, the velocity of the shockwave is greater in the case of the bare Ta. Fig.\ref{fig:Actual}(d)-(e) stem from the nanosecond pump. Similarly the bare Ta produces a faster shockwave. Peak intensity used was 1.69 $\times$ 10\textsuperscript{9}~W/cm\textsuperscript{2} on bare and 1.59 $\times$ 10\textsuperscript{9}~W/cm\textsuperscript{2} on plastic-coated Ta. Relative to the femtosecond pump, the shockwave begins propagation around 10~ns after the peak of the pump.

\begin{figure}[hbt!]
\centering
\fbox{\includegraphics[height=.5\textwidth]{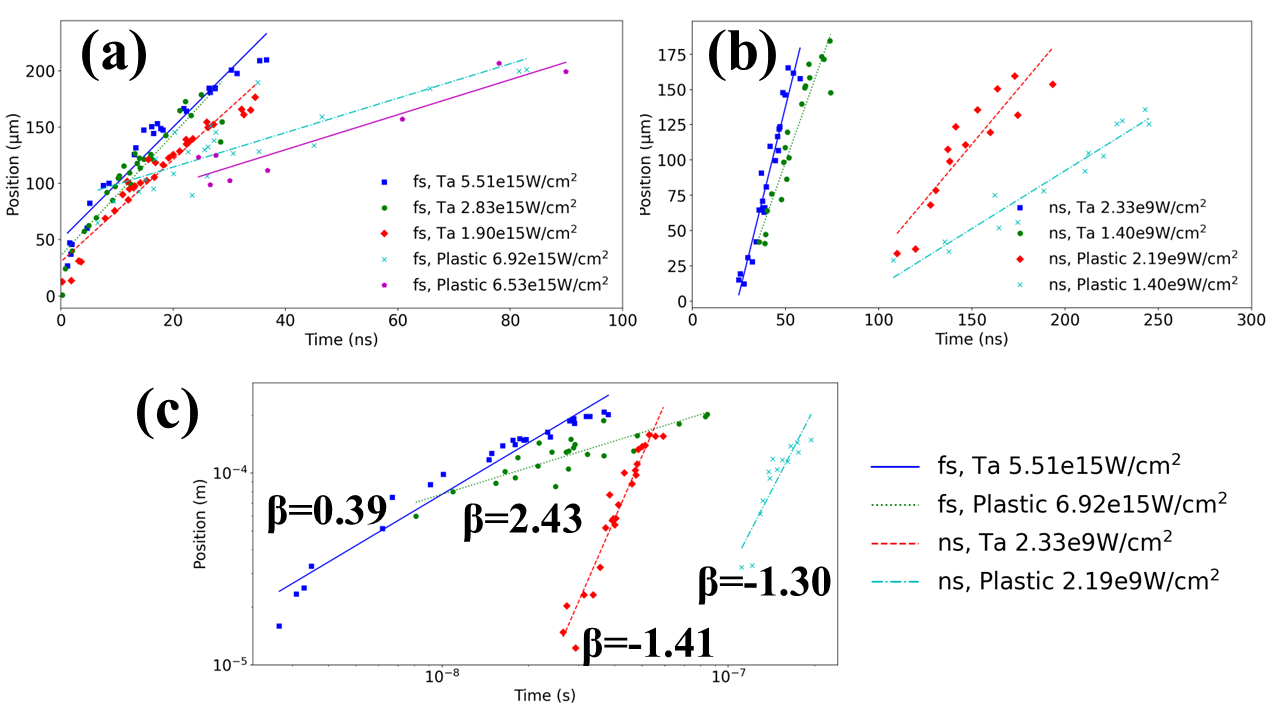}}
\caption{(a)-(b) Shockwave front distance versus time graph for the tested intensities for the 40~fs pump and the 20.4~ns pump respectively. (c) Linear fitting for the highest intensity on each target whilst plotting log(time) and log(distance).}
\label{fig:Shockwaves}
\end{figure}

The shockwave front position within the first 250~ns are plotted in Fig. \ref{fig:Shockwaves}. Fig. \ref{fig:Shockwaves}(a) is obtained from the femtosecond pump, and (b) is obtained from the nanosecond pump. Linear fits are applied such that the velocity can be easily calculated. Additionally, a log-log plot of distance and time is used in Fig. \ref{fig:Shockwaves}(c) for fitting of the Sedov-Taylor scaling \cite{sedov1993similarity}:

\begin{equation}
  R=\lambda \left(\frac{E}{\rho} \right) ^{1/(\beta+2)} t_p^{2/(\beta+2)}
\end{equation}

In this equation, \textit{R} is the radial or longitudinal expansion distance, $\lambda$ is a dimensionless constant equal to unity, \textit{E} is the energy released from the shock, $\rho$ is the air density, \textit{t\textsubscript{p}} is the time of the probe, and $\beta$ is the dimensionality of propagation. When $\beta$=1 the shockwave is a planar wave, $\beta$=2 is cylindrical, and $\beta$=3 represents a spherical emission. By taking the logarithm on either side, is it possible to determine the dimensionality of the shockwave. In Fig.\ref{fig:Shockwaves} only the femtosecond pump fits the model. Due to the nature of this imaging, only the expansion of the shockwave in the longitudinal direction is visible. 

\subsection{Femtosecond pump on borosilicate glass coverslip}
\begin{figure}[h]
\centering
\fbox{\includegraphics[height=.45\textwidth]{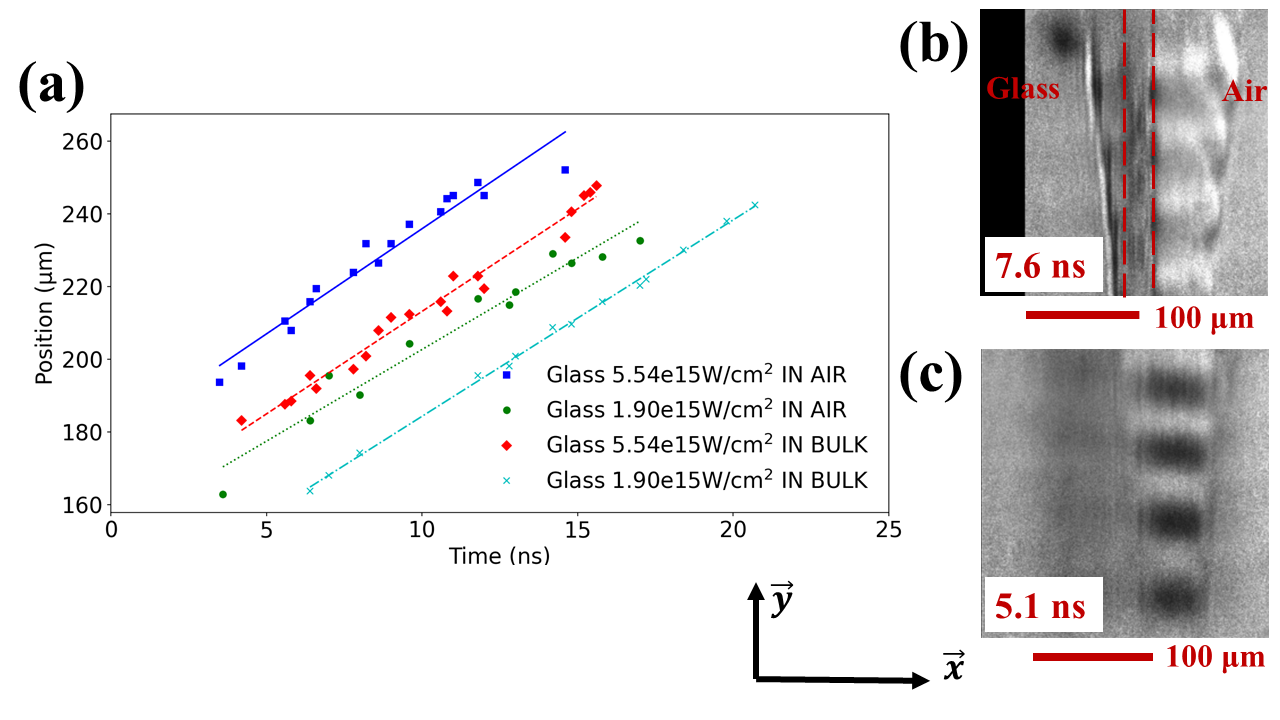}}
\caption{(a) Position of shockwave front versus time for the femtosecond pump on borosilicate, in the transverse direction. (b) The pump with I\textsubscript{p}=5.54 $\times$ 10\textsuperscript{15}~W/cm\textsuperscript{2} is incident from the right side of the image, and propagates in -$\protect\vv{x}$. The dashed area is the air/glass interface. (c) Transmission probe image of the same target, radial expansion of the plasma and shockwave is visible. Peak intensity on target is 3.60 $\times$ 10\textsuperscript{15}~W/cm\textsuperscript{2}.}
\label{fig:Shockwaves Glass}
\end{figure}

The 780~nm, 40~fs pulse was also incident on a borosilicate glass microscope coverslip. This was performed in the same transverse setup as the tantalum targets, but also in transmission with 45\textdegree~angle of incidence. In Fig. \ref{fig:Shockwaves Glass} the shockwave position over time is plotted (a), along with representative transverse and transmission probe images (b)-(c) respectively. In the transverse configuration, the edge of the borosilicate was damaged, whereas the front face was damaged for transmission images. In Fig. \ref{fig:Shockwaves Glass}(b), the output velocities are relatively similar between shockwaves in air or in the bulk. Plasma is clearly visible in the first 20~ns in the transmission setup (Fig.~\ref{fig:Shockwaves Glass}(c)) but the shockwave quickly dissipates and exits the field of view.

\section{Discussion}
\subsection{Effectiveness of Imaging Scheme}

\begin{figure}[hbt]
\centering
\fbox{\includegraphics[height=.4\textwidth]{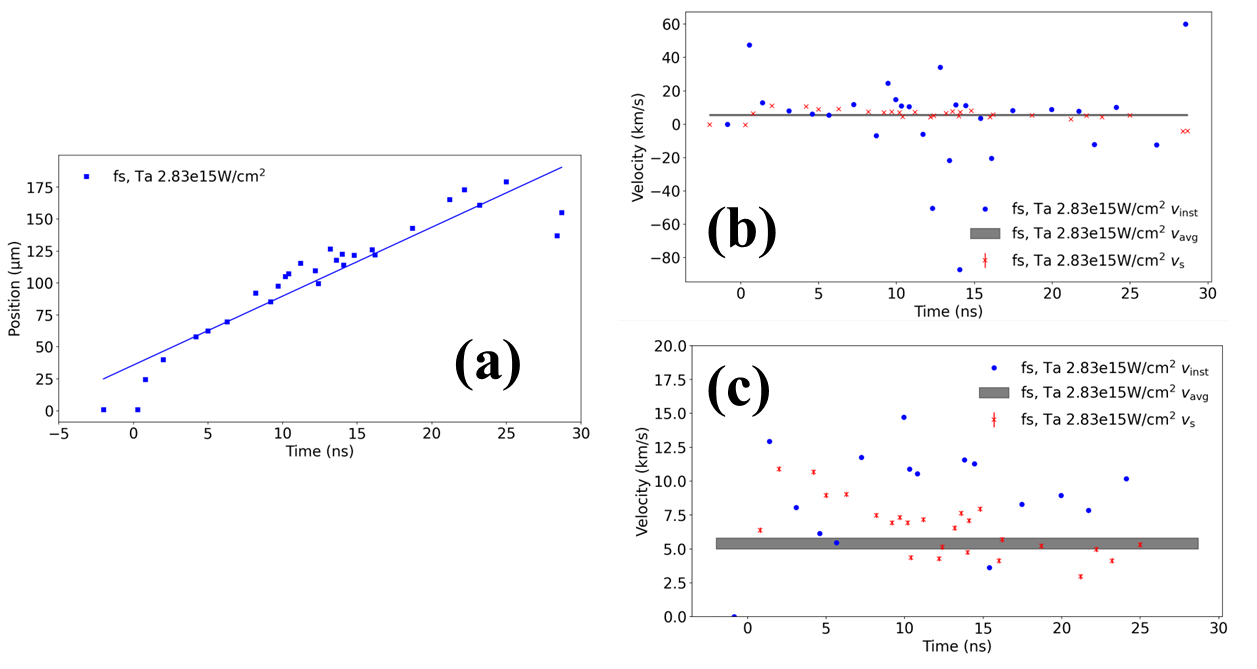}}
\caption{Demonstration of the single-imaging velocity measuring technique. (a) Distance versus time with a linear best fit for bare Ta shot with the femtosecond laser at 2.83 $\times$ 10\textsuperscript{15}~W/cm\textsuperscript{2}, using the 530.57 nm channel (see Fig. \ref{fig:Spectrum}). Each point is determined from an individual pump shot. (b) Velocity vs time graph, with velocities obtained in two ways: first, from graph (a) plotted in blue squares, along with those obtained from multi-time frame probe shots for each time instants, represented by red crosses. The slope of the distance versus time graph is the average multi-image velocity, or \textit{v\textsubscript{avg}}. (c) Zoomed in version of plot (b), standard deviation of $v_{avg}$ and $v_s$ is shown.}
\label{fig:single}
\end{figure}

The pulse train probe works as intended and is detailed in Fig. \ref{fig:single}. In Fig. \ref{fig:single}(a) the position versus time is plotted for the femtosecond pulse on bare tantalum, the velocities obtained in (b) are derived using several different methods. The slope of the linear fit in (a) \textit{v\textsubscript{avg}} is the average multi-image velocity in (b). Instantaneous multi-image velocity, or \textit{v\textsubscript{inst}}, is the slope between two adjacent points on the distance versus time graph. Finally, the single-image velocity \textit{v\textsubscript{s}} is obtained using the multiple channels of the probe.

From Fig. \ref{fig:Spectrum}(b), channel one is the top horizontal band (530.53~nm), while channel four is the bottom horizontal band (533.43~nm). It is possible to obtain the velocity of the shockwave in a single pump shot using this equation:

\begin{equation}
  v_{s}=\frac{x_{ch1} - x_{ch4}}{t_{ch1} - t_{ch4}}
\end{equation}

Where  \textit{v\textsubscript{s}} is the single-image velocity in Fig. \ref{fig:single}(b),  \textit{x} is the position of the shock front, and  \textit{t} is the relative timing between the channels. Channel one and four have a relative time difference of 3.5~ns. The single-image velocity in Fig. \ref{fig:single}(b) is much more similar to the average shockwave velocity, than the instantaneous multi-image velocity. The small deviation between \textit{v\textsubscript{s}} and \textit{v\textsubscript{avg}} may be due to the line of best fit in Fig. \ref{fig:single}(a), which only has R\textsuperscript{2}=0.862, so  \textit{v\textsubscript{s}} may be more accurate than \textit{v\textsubscript{avg}}. Individual differences between pump shots in terms of energy or pointing can be offset by using this single-image velocity measurement technique. 

This image is distorted vertically but is unaffected horizontally. More channels could be imaged on the camera as seven were present, but low signal made this unfeasible. The presence of more or fewer channels can be altered by moving the grating closer or further to the camera respectively. In addition, as each channel is separately delayed by a fiber, changing the relative timing of the pulse train would be trivial. Another advantage of this imaging scheme is the ease in removing the pump light, as the imaging setup features a diffractive element.

Fringe diffraction patterns are present in some images, Fig. 
\ref{fig:Shockwaves}(d), due to minute alignment changes in the sample, and could mostly be corrected. Finally the 527~nm pump-flash was difficult to mitigate at the higher intensities. While 527~nm was out of the viewing window of the camera, plasma emission centered at this wavelength leaked into timed probe images.

\subsection{Bare Ta and plastic on Ta}

\bigskip
{
\centering
\begin{tabular}{ |p{2cm}||p{2cm}|p{2cm}|p{2cm}|p{1.5cm}|p{1cm}|  }
 \hline
 \multicolumn{6}{|c|}{Table 1: Calculated Shockwave Velocities} \\
 \hline
 Target&I\textsubscript{p} (W/cm\textsuperscript{2})&Wavelength (m)&Pulse Duration (s)&Velocity (km/s)&R\textsuperscript{2}\\
 \hline
 Ta   & 5.51 $\times$ 10\textsuperscript{15}   &780 $\times$ 10\textsuperscript{-9}&   40 $\times$ 10\textsuperscript{-15}&5.01&   0.939\\
 & 2.83 $\times$ 10\textsuperscript{15}   &780 $\times$ 10\textsuperscript{-9}&   40 $\times$ 10\textsuperscript{-15}&5.39&   0.862\\
 & 1.90 $\times$ 10\textsuperscript{15}   &780 $\times$ 10\textsuperscript{-9}&   40 $\times$ 10\textsuperscript{-15}&4.54&   0.937\\\cline{2-6}
 & 2.33 $\times$ 10\textsuperscript{9}   &527 $\times$ 10\textsuperscript{-9}&   20.4 $\times$ 10\textsuperscript{-9}&5.32&   0.939\\
 & 1.40 $\times$ 10\textsuperscript{9}   &527 $\times$ 10\textsuperscript{-9}&   20.4 $\times$ 10\textsuperscript{-9}&3.75&   0.901\\
 \hline
 Plastic on Ta   & 6.92 $\times$ 10\textsuperscript{15}   &780 $\times$ 10\textsuperscript{-9}&   40 $\times$ 10\textsuperscript{-15}&1.53&   0.721\\
 & 6.53 $\times$ 10\textsuperscript{15}   &780 $\times$ 10\textsuperscript{-9}&   40 $\times$ 10\textsuperscript{-15}&1.56&   0.894\\\cline{2-6}
 & 2.19 $\times$ 10\textsuperscript{9}   &527 x 10\textsuperscript{-9}&   20.4 $\times$ 10\textsuperscript{-9}&1.58&   0.818\\
 & 1.40 $\times$ 10\textsuperscript{9}   &527 $\times$ 10\textsuperscript{-9}&   20.4 $\times$ 10\textsuperscript{-9}&0.82&   0.933\\
 \hline
 Borosilicate glass, Air   & 5.54 $\times$ 10\textsuperscript{15}   &780 $\times$ 10\textsuperscript{-9}&   40 $\times$ 10\textsuperscript{-15}&5.79&   0.948\\
 & 1.90 x 10\textsuperscript{15}   &780 $\times$ 10\textsuperscript{-9}&   40 $\times$ 10\textsuperscript{-15}&5.04&   0.954\\
 \hline
 Borosilicate glass, Bulk   & 5.54 $\times$ 10\textsuperscript{15}   &780 $\times$ 10\textsuperscript{-9}&   40 $\times$ 10\textsuperscript{-15}&5.64&   0.978\\
 & 1.90 $\times$ 10\textsuperscript{15}   &780 $\times$ 10\textsuperscript{-9}&   40 $\times$ 10\textsuperscript{-15}&5.40&   0.998\\
 \hline
\end{tabular}\par
}
\bigskip
Average shockwave velocities obtained from linear fitting of the plotted distance versus time, for all trials, is detailed in Table 1.

It generally follows from Table. 1 that higher energies lead to higher shockwave velocities which agrees with \cite{liu2024ultrafast}. Sedov scaling generally fits with the femtosecond pump on bare and plastic-coated tantalum. In Fig. \ref{fig:Shockwaves}, for bare Ta, a $\beta$ around 1 indicates a strong planar wave, which may be due to the plasma \cite{zhang2023ultrafast}. For the plastic-coated Ta, a $\beta$ of 2.43 may indicate that the shockwave begins as cylindrical and becomes spherical \cite{zhang2023ultrafast}. Two shockwaves were generated from 3.83 x 10\textsuperscript{14}~W/cm\textsuperscript{2} off the surface of alumina, the first of which was propagating at 20~km/s at 800~ps, and the second was propagating at 10~km/s after 800~ps \cite{zhang2015influence}. The second shockwave measured had a $\beta$ = 2, or a cylindrical symmetry. In our case, the shockwave generated from the bare tantalum is likely an analogue to their first shockwave generated, with its higher velocity (due to a plasma channel) \cite{zhang2015influence}. 

The nanosecond pump generated shockwaves of 0.82 - 5.32~km/s, which corroborates with literature values, going from 7.1~km/s to 1~km/s at 1~ns \cite{liu2024ultrafast}. Exponential decay of shockwave speed, agreeing with the Sedov-Taylor equation is observed by Liu et al. when damaging fused silica with a 355~nm, 6.65 x 10\textsuperscript{10}~W/cm\textsuperscript{2}, 4~ns pulse duration \cite{liu2024ultrafast}. They observe shockwaves begin tens of nanoseconds after the pump interaction, which agrees with our bare Ta trial; whereas, our measured shockwave off of the plastic-coated target may in fact be ejected material. After $\sim$50~ns Liu et al. observe material splashing off the front of the target, traveling between 0.15-2.0~km/s, which may align with our trial on plastic-coated Ta.

Sedov scaling does not fit well with our measured nanosecond-generated shockwaves. The nanosecond damage may not agree with the Sedov scaling due to a large beam diameter, 178.0~$\mu$m, which is assumed to be a point source in the model \cite{zel1967physics,nagel2021single}. Jeong et al. observe that the shockwave more closely follows Sedov scaling as the spot size is reduced \cite{jeong1998propagation}. Additionally, shockwaves produced from laser ablation propagate faster than the model predicts on a short time scale after the pump interaction \cite{palanco2014particle}, which may be due to the generation and expansion of the plasma.

\subsection{Femtosecond pump on borosilicate glass coverslip}
The shockwave in the bulk of borosilicate glass, in Table. 1, aligns well with literature values for which the speed of sound is 5.56 - 5.69~km/s \cite{chocron2016damage,orphal2009failure}. Previous literature also generates a pressure wave at the speed of sound in glass for a femtosecond laser \cite{ito2019dynamics} which maintains a constant speed; however, their shockwave in air does follow the Sedov-Taylor equation but it may be due to a lower intensity with a longer pulse duration (2.66 x 10\textsuperscript{14}~W/cm\textsuperscript{2} at 210~fs), in addition they observed the shockwave over 300~ns. The pressure wave in air travels faster than that in the bulk until around 5~ns \cite{ito2019dynamics}, but the shockwave in air is always supersonic. In silica glass at 7.52 x 10\textsuperscript{14}~W/cm\textsuperscript{2}, shockwaves were observed in the bulk decreasing from 8.7 to 6.0~km/s (for the first 16~ns) where 5.76~km/s was the speed of sound \cite{hu2010generation}; additionally, subsequent shockwaves were observed in the bulk from 11~ns after the pump interaction. While no secondary or tertiary shockwaves were visible in this experiment in the glass, Hu et al. describe the magnitude decreasing with each subsequent pressure wave \cite{hu2010generation}.

\section{Conclusion}
Laser plasma interaction dynamics from both a femtosecond and nanosecond pump, were captured via a novel probe scheme showing varying shockwave geometries and velocities in both air and inside transparent amorphous samples. The ultrashort pump interaction generated supersonic shockwaves expanding in air from bare and plastic/polymer coated Ta agrees well with the literature. The shockwave from bare Ta exhibits planar geometry due to a plasma channel, whereas the plastic-coated Ta followed cylindrical and spherical Sedov-scaling, with a significantly slower shockwave speed. 

The nanosecond pump driven shockwave evolution does not agree well with Sedov scaling, but is likely due to a large focal spot size \cite{jeong1998propagation}. Images of the femtosecond pump in borosilicate glass illustrate a pressure wave at the speed of sound in the bulk, and a shockwave emission in air. Both pressure and shockwaves in bulk and in air agree with literature values. 

The imaging scheme works as intended, and one can obtain the velocity of the shockwave from a single pump shot reliably. The single shot imaging agrees well with the average multi-image velocity. This pulse train is flexible in its number of visible channels, as well as its relative timing between probes and field of view, which could be as short as 10 ps. Wavelengths for each channel can also be separately tuned. Although currently the probe pulse train is in the visible wavelength range (from second harmonic generation of the near-IR fundamental wavelength), a UV pulse train using third or fourth harmonic generation is possible, to probe denser plasma dynamics. Future work will improve the design of the Spectral Energies probe laser to increase the energy of the pulses and increase temporal tunability.

\begin{backmatter}
\bmsection{Funding}
The authors acknowledge funding support from  the DOE SBIR/STTR program under the Fusion Energy Science (FES) program with contract no. DE-SC0022743.

\bmsection{Acknowledgment}
The Authors acknowledge help from Mr. Alex Blackston for coating some Ta samples with polymer/plastic.

\bmsection{Disclosures}
The authors declare no conflicts of interest.

\bmsection{Data Availability Statement}
Data underlying the results presented in this paper are not publicly available at this time but may be obtained from the authors upon reasonable request.

\bmsection{Supplemental document}
See Supplement 1 for supporting content.

\textbf{If you use supplement, be sure to refer to it in the main body.}

\end{backmatter}

\label{sec:refs}

\bibliography{Bibliography}
\end{document}